# Observing instantons directly on the lattice without cooling

Eduardo Mendel [1] and Guido Nolte [2]

*FB Physik, Carl von Ossietzky Universität Oldenburg,*
*26111 Oldenburg, Germany*

November 29, 1995

**Abstract**

Based on the study of the simple Abelian Higgs model in $1+1$ dimensions we will present a new method to identify and localize extended instantons. The idea is to measure the topological charge on regions somewhat larger than the extended instantons so as to average out the ultraviolet fluctuations but without losing the detailed topological information when going to the full space. The instanton size and probability density can be directly extracted from this analysis. Local dislocations, which can be avoided for fine enough lattices, can be reinterpreted as modified boundary conditions producing sectors with net topological charge.

## 1 Introduction

Topological non trivial configurations, such as instantons [1] are expected to play an important role in such diverse phenomena as net Baryon number generation [2, 3, 4], the $U(1)$ problem [2, 5] or even quark confinement [6, 7]. After the first attempts to calculate topological effects with semiclassical methods [8, 9], where it is very hard to find the relative weight for the various configurations or to go beyond the dilute gas approximation, several works have tried to extract directly the topological content from lattice simulations [10, 11, 12, 13]. Up to the present, most simulations have been done on fairly coarse lattices where the instanton sizes are of the order of the spacing, thus making it hard to distinguish between a physical instanton and a lattice artifact (dislocation). Furthermore, the topological charge is usually only measured on the whole volume where, as a consequence of the periodic boundary conditions, we can only see a net charge if we have dislocations (which per definition are points where the charge gets shifted). In several works cooling [15, 14] and improved actions [17, 16] are being used in order to smoothen out the instantons and filter the physical ones, leaving some uncertainty on the agreement of the topological content with the one in the starting model.

In this work we want to present an alternative method which extracts directly the topological information from the lattice configurations. It consists in measuring the topological charge on volumes larger than or of the order of the relevant instantons. Let us first assume that one can reach a coupling regime where the gauge configurations are smooth enough so that there are no individual sites where the topological density gets even close to $\pm 1/2$. Then one has safely configurations without dislocations and if we take periodic boundary conditions the total geometrical topological charge is exactly zero. This does not mean that there are no instantons in the system, but rather that there is a number of instantons and anti-instantons with total charge zero. Furthermore, in the thermodynamic limit of very large volumes $V$, the topological charge probability distribution in some finite sub-volume $v$ should reach a fixed limit $P_v(Q)$.

---

[1] Dedicated to the memory of my beloved brother, Prof. Roberto Mendel
[2] New Address: Klinikum Steglitz, Berlin

This probability distribution to get some charge $Q$ in a volume $v$, which will be obtained from the lattice simulation, should have a broad envelope for very large volumes, which scales like $P_{\alpha v}(Q) = 1/\sqrt{\alpha} P_v(Q/\sqrt{\alpha})$ with $\alpha v$, with distributions centered at integer $Q$'s reaching the envelope. As we reduce the volume $v$ to sizes smaller than the instanton, we see only a distribution peaked at $Q = 0$ as we have only pieces of instantons in $v$ or fluctuations. In fact, for very small volumes one sees only the ultraviolet fluctuations in the topological charge giving a narrow Gaussian distribution ($\sigma \ll 1$). Interestingly, these are topological fluctuations in regions of stable topological vacuua and therefore tend to cancel out over fairly short distances. Due to their topological character the total dispersion in the charge that they produce only grows as $\sqrt{s}$ with $s$ being the free surface. For larger volumes, of the order or somewhat larger than the instantons, we start getting relevant contributions to $P_v(Q)$ from instantons, where the charge adds coherently to $\pm 1$ (if they are fully in $v$), thus broadening clearly the charge distribution in contrast to the fluctuations. Note that instantons extremize locally the action and therefore should not be considered as fluctuations themselves. At even larger volumes one can do a careful analysis using Poisson distributions of almost free multi-(anti)instantons with some average density $\rho$ and instanton size $v_I$, in order to determine these parameters for a given simulation. Then we can check the scaling with $v$ as described above. The general idea is therefore to look at volume scales $v$, where

$$a^D \ll v_I < v \ll V$$

so that we have mainly supressed the fluctuations and can study the ensemble of (anti)instantons, while having effectively almost free spatial boundary conditions. We have tested these ideas for the abelian Higgs model in $1 + 1$ dimensions where one has a simple system with non-trivial topology. We find indeed, for couplings corresponding to fine lattices, clearly identifiable extended instantons over several lattice spacings, while the total charge over the whole $V$ stays always zero. Assuming an almost free ensemble of (anti)instantons, we have been able to extract an instanton average density $\rho$ and size, also as function of temperature, $T = 1/N_\tau$. The density $\rho$ which can also be easily extracted from the second moment of $P_v(Q)$ for large $v$, corresponding to the topological susceptibility, falls steeply at higher $T$ as the instantons get squeezed in the time direction [24]. We have also studied the evolution of the position of the instantons in Monte Carlo time and found that they get created locally as instanton - antiinstanton pairs and annihilate again after some time or drift away, to get destroyed later when they encounter another pairing.

As for some models it might get prohibitively expensive to go to such fine lattices where the topological density does not get close to $\pm 1/2$ for some sites (it might even be impossible for cases where the instantons stay almost scale invariant after renormalization) we wanted to analyze carefully what happens when at some individual plaquettes (hypercubes in higher D ) the geometrical definition for the topological density surpasses $\pm 1/2$ before taking the "mod" operation, which brings it to the interval $(-1/2, 1/2]$. At those points where one had to make the "mod" operation, one introduces "defects" to the topological configuration in such a way that the total charge is not zero anymore. We will come to the conclusion, for the abelian Higgs model, that these defects only make a difference when there is a net number of them as if we had new boundary conditions with this number of instantons, otherwise their effect cancels out. In cases with net number $N$ of defects in a total volume $v$ the remaining configuration behaves just as if we were in a case without singularities in a sub-volume $v$ with exactly $N$ instantons. Furthermore, if we count in Monte Carlo time with which recurrence the configuration stays in each topological sector we find a probability distribution very similar to the $P_v(Q)$ one without singularities.

In our discussion up to this point we have taken the geometrical definition for the topological density which always gives 0 (or an integer for dislocations) in a periodic box. This geometrical definition gets fairly cumbersome in non-abelian theories in higher D [19, 20] but should still give integer results and our whole idea should work there too. We have also investigated the field theoretic definition for the topological density which is easier to implement [21] and gives the right naive continuum limit but does have only approximately right addition properties and therefore does not give exactly integers on the whole volume $V$. Nonetheless, on volumes $v$ somewhat larger than the instantons as we only need for our analysis, and for not too coarse lattices, the field theoretic definition gives good agreement with the geometrical one.

In the next section we will discuss the abelian Higgs model in $1 + 1$ D and its topological density definitions. In section three we will choose the parameters in the model in order to have effectively a fine lattice. In the fourth section we will identify the instantons in our simulations and present a simple model that fits the topological charge distribution. In the fifth section we will discuss the role of dislocations and the field theoretic definition for the topological charge density. Finally we will discuss our results and propose the use of this method in the non-abelian case.

## 2  The Model and defining a Topological Density

Starting point of our analysis is the lattice formulation of the abelian Higgs model in 1+1 dimension with the Euclidean action

$$S = -\beta \sum_n Re(U_n) + \lambda \sum_n ((\Phi_n^*\Phi_n - 1))^2 - 2\kappa \sum_{n,\mu} Re(\Phi_n^* U_{n,\mu} \Phi_{n+\mu}) \ . \quad (1)$$

Here $\Phi_n$ is the complex Higgs field, $U_{n,\mu} = \exp(i\theta_{n,\mu})$ are the usual links and $U_n = \Pi_\square U_{n,\mu}$ the plaquette variables. (Note that in two dimensions the plaquette does not need indices.) This is one of the simplest models with non-trivial topological content.

In order to define univocally the action in the lattice simulation, we only need the $U_{n,\mu}$ gauge fields and do not need to keep track of the phases $\theta_{n,\mu}$. To get the topological information we could think naively of extracting the $\vartheta_n$ phases from the fundamental plaquettes as

$$\exp(i\vartheta_n) := U_n \ . \quad (2)$$

which is ill defined. The ambiguity in the definition of these $\vartheta_n$ would give rise to the existence of artificial instanton configurations. We will see shortly that if we keep the phases $\theta_{n,\mu}$, a more careful definition is possible.

In the continuum theory the topological charge in a region $v$ may be computed easily from the field strength tensor $F_{\mu\nu}$ via

$$Q_v = \frac{1}{4\pi} \int_v d^2x \; \epsilon_{\mu\nu} F_{\mu\nu} \; . \tag{3}$$

Note that it is usually understood to take $v$ to be the whole space-time in which case, assuming a finite action, one gets an integer winding number. It is not an integer if one only assumes a finite action density, so there is no difference from the lattice case with free boundary conditions.

The charge $Q_v$ may be written as an integral over the boundary of $v$

$$Q_v = \frac{1}{2\pi} \int_{\partial v} ds \; n_\mu A_\mu \; , \tag{4}$$

with $n_\mu$ being the normalized tangent vector of $\partial v$. From eq. (4) it is easily seen that for periodic boundary conditions in space and time the topological charge of the whole space-time will always be zero.

There are several possibilities in order to define the lattice version of the topological charge density and from it $Q_v$. The field theoretical definition for $Q_v$, based on the fact that an expansion of the plaquette to lowest order in $a$ gives the field strength[21], is

$$Q_v^F := \frac{1}{2\pi} \sum_{n \in v} Im(U_n), \tag{5}$$

and is clearly free of any ambiguities as it involves the plaquette in a well defined manner. The disadvantage of this definition is that eq.(4) gets spoiled for finite lattice spacings. Notice that this comes from the fact that one does *not* have strictly a *density* in the sense that if we take double the lattice spacing we do not get exactly the same $Q$. Therefore, finite action configurations would not have exactly integer (or zero, for periodic boundary conditions) winding number, but it can get closer for finer lattices.

We can avoid this problem by defining a geometrical charge density [18, 14], which is the simplest example of the fibre bundle construction used in the non-abelian case [19, 10, 20], which is in this case just

$$Q_v^G := \frac{1}{2\pi i} \sum_{n \in v} \log(U_n). \tag{6}$$

This $Q_v^G$ may violate eq.(4) by an integer amount in some cases. In particular you may change the winding number by a continuous and local change in the fields. Take for example a periodic lattice ($Q = 0$) with charge on one plaquette close to $1/2$, so that if one changes slightly one link, the charge on that plaquette will jump to $-1/2$, the remaining plaquettes staying almost the same, thus obtaining $Q = -1$ for the whole lattice. These lattice artifacts can be avoided for fine enough lattices as we will show.

In order to keep track of these ambiguities we will directly work with the phases $\theta_{n,\mu}$ of the links $U_{n,\mu}$ as the fundamental variables, from which $\theta_n$ is given in obvious notation as the sum $\theta_n = \theta_{n,\hat{x}} + \theta_{n+\hat{x},\hat{\tau}} - \theta_{n+\hat{\tau},\hat{x}} - \theta_{n,\hat{\tau}}$. We then define $Q_v$ to be

$$Q_v := \frac{1}{2\pi} \sum_{n \in v} [\theta_n] \tag{7}$$

where the brackets shift $\theta_n$ by an integer multiple of $2\pi$ into the interval $(-\pi, \pi]$. This definition is clearly equivalent to (6), however, using $\theta_n$ as the fundamental variable we are now able to locate lattice artifacts (called dislocations), namely places for which $|\theta_n| > \pi$. Note that in our Metropolis algorithm we only update the $\theta_{n,\mu}$ by small changes from the old values, otherwise we could trivially generate dislocations by adding $2\pi$ to a link, in a plaquette with already low action. (These trivial dislocations come as neighboring pairs which do not affect the physics as we will see in sec. 5.1 ). The experience has shown us that if we choose parameters corresponding to a fine enough lattice, so that the physical instantons are much larger than $a$, we can avoid obtaining any dislocations at all, over the whole Monte Carlo run. It would be interesting to know for the non-abelian case, where a characteristic instanton size is only obtained due to renormalizaton, if one could also avoid completely dislocations.

In any case, for non-abelian gauge theories, the geometrical approach to topological charge is far more involved, and reasonable criteria to avoid (or exploit) the existence of dislocations would be useful, even in order to be able to use safely the simpler field theoretical definition. To this end we will consider in our model both definitions for $Q_v$ and will show in sec. 5.2 that we arrive essentially at the same results. Non-existence of artifacts will be made (extremely) plausible for the field theoretical definition and will be controlled by looking at the full $\theta_n$ before taking the "mod" for the geometrical definition.

In general an instanton is an extended object and in order to identify this object one has to compute the topological charge in a certain region in space-time. This region has to be large enough to contain at least one instanton and it should not be larger than half the full lattice. Analyzing for example the topological charge of the whole lattice would give a nontrivial result only in the presence of dislocations.

An alternative used by many authors is to identify topological excitations from the local topological density. Usually short range fluctuations are much larger than the signal and the noise has to be removed by (non-linear) cooling procedures, which affect the excitations too, probably giving misleading results.

One of the physically most relevant quantities to extract is the mean density of instantons, $\rho$. This can be obtained easily, as we will see, from the charge susceptibility $<Q_v^2>$ on finite volumes $v$, while for the whole $V$ it is again 0 without dislocations.

## 3  Choice of lattice parameters

In order to be able to extract physical information relevant in the continuum from lattice simulations we need that all relevant correlation lenghts be much larger than the lattice cutoff $a$. As for this 2-D model we do not expect a second order phase transition at finite parameters, we can at best hope to find a rapid crossover parameter region where the correlations are long range, in order to identify extended topological objects without having lattice artifacts. Once such a parameter region has been found, we have effectively a fine lattice spacing $a$ and we can make sure that all values of $\theta_n$ stay small during the whole simulation, without producing any dislocations.

With this aim we analyze the correlation functions of the following three operators [22] on a lattice of size $L_x \times L_\tau$ with spatial and temporal coordinates $x$ and $\tau$ respectively.

$$O_s(\tau) = Re\left(\sum_{x,\mu} \Phi^*_{(x,\tau)} U_{(x,\tau),\mu} \Phi_{(x,\tau)+\mu}\right) \quad (8)$$

$$O_v(\tau) = Im\left(\sum_{x,\mu} \Phi^*_{(x,\tau)} U_{(x,\tau),\mu} \Phi_{(x,\tau)+\mu}\right) \quad (9)$$

$$O_\gamma^k(\tau) = \sum_x \exp(ikx) Im(U_{x,\tau}) \quad (10)$$

corresponding to the scalar and vector mesons and the photon with momentum $k_j = 2\pi j/L_x$. We have to take the connected correlation function in cases where there is a non-zero vacuum expectation value $< O_i >$. Note that for the $\gamma$ we consider higher momentum states which couple to the real photon [23], while the $O_\gamma^{k=0}$ operator which corresponds to the field theoretic definition for the topological charge in 2-D, gives us almost zero gap energy corresponding to the almost degenerate vacuum states which are accessible in the presence of instantons. Even if such a picture seems to hold qualitatively ( the correlation of $O_\gamma^{k=0}$ reaches zero mass when the coupling $\kappa$ gets low enough so as to start producing instantons), it would be interesting to study this more carefully in the future.

We are not going to give a detailed discussion of the phase structure. Instead we merely state that there is a rapid crossover, between regions similar to the Coulomb and Higgs phases in higher D. The masses $m_s$, $m_v$ and $m_\gamma(k \neq 0)$ all have a minimum in this region while $m_\gamma(k = 0)$ which is almost degenerate with $m_v$ and $m_\gamma(k \neq 0)$ (right dispersion relation here) at high $\kappa$, drops to zero when we lower $\kappa$ across the region and below. The crossover roughly takes place at a set of parameter pairs $(\lambda, \kappa)$ laying on a straight line between the points $(.1, .35)$ and $(.5, .45)$. This line of minima is not very sensitive to the gauge coupling $\beta$. As in this paper we are interested in a very clear distinction between lattice artifacts and extended (physical) instantons, a large value of $\beta = 10.$ has been chosen, corresponding to very fine lattices where dislocations practically do not appear anymore. In sec.5, for the purpose of studying the effect of dislocations, we will reduce the gauge coupling to $\beta = 6.3$. Most simulations have been done then around the "scaling region", at $(\lambda, \kappa) = (.2, .37)$, for which we have obtained the masses, $m_s \approx .65$, $m_v \approx .57$, $m_\gamma(k \neq 0) \approx .60$ and $m_\gamma(k = 0) \approx .02$ in lattice units. We are in the region where this $m_\gamma(k = 0)$ drops almost to zero signalling the appearance of instantons as we lower $\kappa$. We will see in the next section that with these parameters we get fairly extended instantons with a sizable topological mean density.

## 4 Identifying Instantons

For a clear isolation of instantons it is useful to consider first the case of small instanton density, which can be obtained by raising the temperature [24] to high T. Thus we will choose narrow lattices in the temporal direction. On the other hand being close to the continuum limit implies large instantons in lattice units, and if we want to allow the system to create instanton-antiinstanton pairs we need to have large lattices in spatial direction. Consequently we will use extremely non-quadratic lattices typically being of the order of $6 \times 160$.

In general the observed quantity will be the topological charge $Q_v$ in a rectangular volume with size $L_x \times L_\tau$ and with center $(x, \tau)$. However, in the high temperature regime you would not find strong structural dependence in the temporal direction and we will take $L_\tau = N_\tau$ throughout this paper, in which case $Q_v$ doesn't depend on $\tau$. This has the added nice feature of decreasing radically the ultraviolet fluctuations, which grow as the square root of the surface, in this case just $2N_\tau$ ( due to periodicity, the time "boundary" phases cancel). The $L_x$ has to be somewhat large than an instanton and we take values between 20 and 160 lattice spacings.

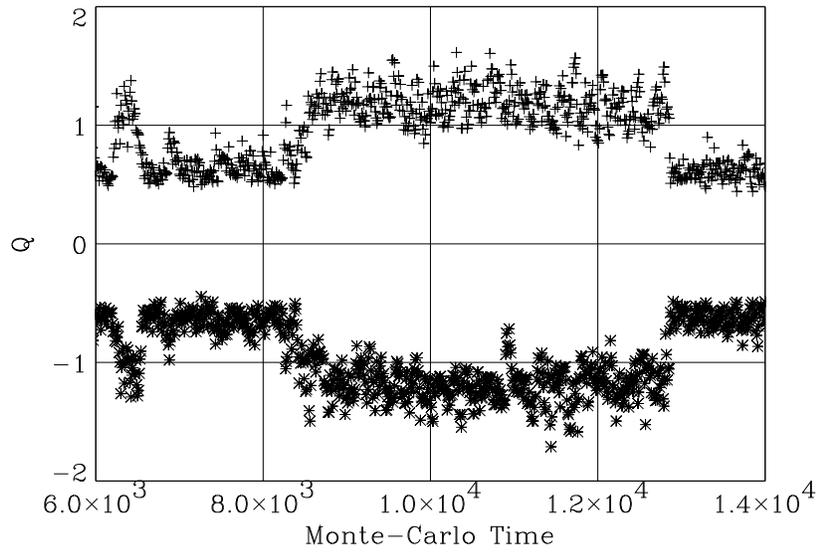

Figure 1: Inst.-antiinst. pairs are created(destroyed) in Monte Carlo time as seen from the sudden rise(fall) in spread in $Q_v$ at some two positions. ($v = 20 \times 4$)

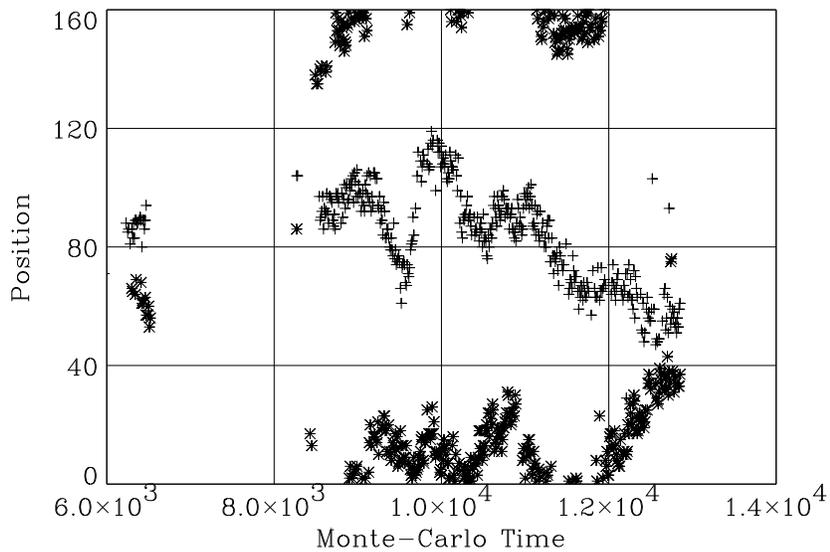

Figure 2: Same run showing the position where the charge $Q_v$ is max.(min.) and greater than 1.0, indicating presence of inst.(antiinst.) there.

We perfom the simulation with a Metropolis algorithm using a multi-hit updating with 10 hits per update with small increments in the phases so as to avoid artificial dislocations. Depending on the observed quantity the number of iterations is in the range of $10^4$ to $10^6$. We want to emphasize that if not otherwise stated the phases $\theta_n$ stay safely small ($\theta_n \ll \pi$) for all positions and iterations. Thus we are free of lattice artifacts.

Doing now a Monte Carlo simulation we observe that the system makes rapid jumps from a vacuum configuration (including fluctuations) and metastable states consisting of instanton-antiinstanton pairs. Depending on the parameters two or more pairs or sometimes even instantons with charge $\pm 2$ are distinguishable.

To demonstrate this we show in Fig.1 for a $4 \times 160$ lattice the maximal (minimal) measured $Q_v$ value for any position $x$, on volumes $v$ ($4 \times 20$). We plot only when $|Q| > 1.0$ and averaged over the last 10 runs ( just for visualization purposes, only for this plot). We observe in this histogram clear jumps for the values of $Q_v$ by almost one unit. We interpret this as creation or annihilation of an instanton-antiinstanton pair with topological charge $\pm 1$. The jump looks indeed to be less than one unit because the probability to have a large positive fluctuation on top of the instanton charge in a volume $v$ containing it, is much smaller than the probability to have the fluctuation somewhere else.

The dynamics of the pair in Monte-Carlo time is visualized in Fig.2. We have plotted for the same simulation the location of the maximal (minimal) $Q_v$ as a function of Monte-Carlo time $T$. For $T \approx 6300$ a short lived pair appears at location $x \approx 80$, then there is a period with no instantons until at $T \approx 8500$ a new pair is created at $x \approx 120$. These excitations move fairly independent and randomly through the lattice, even winding around the lattice and finally annihilate each other at $x \approx 50$. The mean size of the (anti)instantons, $L_I$, can be first estimated crudely, and will be shown to range between 15 and 30 for most of our simulations.

For a quantitative analysis of the data we will evaluate the probability distribution $P_v(Q)$ to get a certain charge $Q$ in a sub-volume $v$ of $V$, which will be discretized into classes of width $\Delta Q = .1$. For pure fluctuations the distribution is expected to be gaussian around $Q = 0$. As we see from Fig.3 and Fig.4 the true distribution is far from being gaussian. Instead we observe clear peaks around integer values of $Q$ caused by the (anti)instanton excitations with integer topological charge. The peaks are not precisely at integer values due to the overlap with gaussian tails caused by the fluctuations and due to the finite size of the instantons, as we will show.

We have fitted the probability distribution of $Q$ with a simple model where we assumed (almost) non-interacting instantons and antiinstantons with topological charge $\pm 1$ and with a uniform charge density on its size $L_I \times N_\tau$.

With $\rho$ being the average density of instantons (or antiinstanstons), which we want ultimately to determine, the probability $p_{n,m}(\rho V)$ to have $n$ instantons and $m$ antiinstantons in the whole

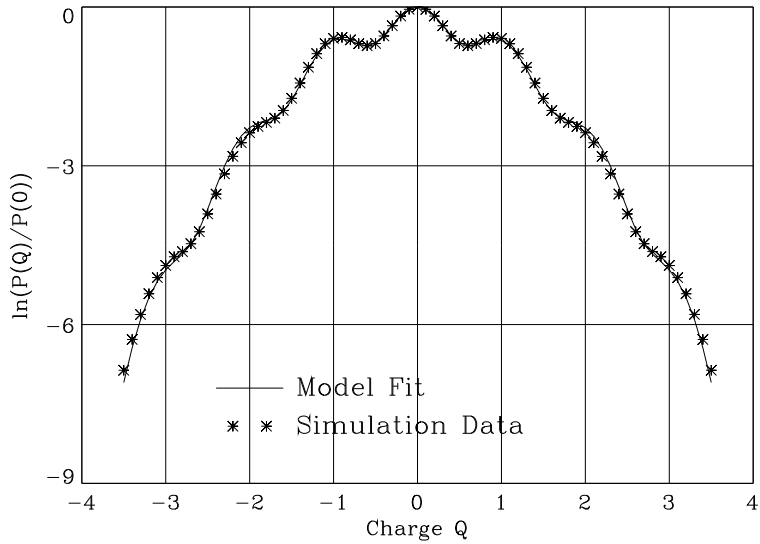

Figure 3: Probability distribution to find a top. charge $Q$ in volumes $v = 160 \times 6$, half of $V$. Note the agreement with the data over 3 orders of magnitude. The enhancements at integers correspond to multi-instantons fully in $v$.

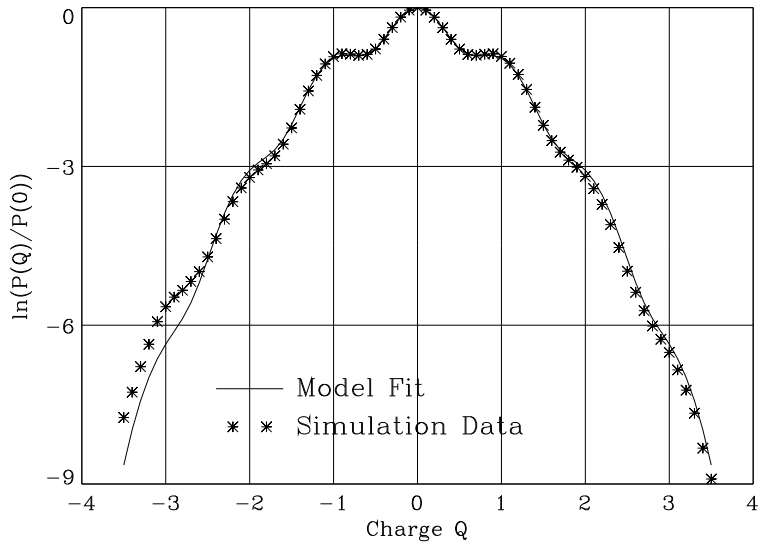

Figure 4: Same as above but with only half the volumes $v$ over which the charge is measured. Now the probability is lower to have multi-instantons in $v$, thus less pronounced peeks than above.

volume $V$ is equal to the product of two poissonian distributions [15]

$$p_{n,m}(\rho V) = \frac{(\rho V)^{n+m}}{n!\, m!} \exp(-2\rho V) \,. \tag{11}$$

Taking into account that the boundary conditions enforce equal numbers of instantons and antiinstantons the probability $p(n)$ to have $n$ pairs is now given by the diagonal elements of (11)

$$p_n(\rho V) = \frac{(\rho V)^{2n}}{n!^2} \frac{1}{N} \,, \tag{12}$$

where $N$ is the normalization.

With $p_n$ being the probability weight to have $n$ pairs, we have to ask now, in how many ways can we get a certain charge $Q$ in the sub-volume $v$, by displacing the $n$ extended instantons and $n$ antiinstantons to all possible locations in $V$. This number of possibilities to get some $Q$, normalized so as to get with certainty any conceivable one in this sector, gives us the charge probability distribution for $n$ pairs, $f_n^v(Q)$. The total charge probability distribution, $P_v(Q)$, is then just

$$P_v(Q) = \sum_n p_n f_n^v(Q) \,. \tag{13}$$

For pointlike instantons, this distribution would clearly only give sharp peaks at integer $Q$'s as the topological charge would be in or out of $v$. For extended ones, the charge can be partially in $v$, smearing the peaks around the integers. One can solve analytically $f_n^v(Q)$ for $n = 1$, (giving peaks at 0 and $\pm 1$ and a plateau in between), but for higher $n$'s one has to solve it numerically.

Up to this point, we have then obtained a charge probability distribution assuming a perfect gas of (anti)instantons with a uniform charge density on their size $v_I$. In reality, we have to consider also the short range fluctuations which fortunately only grow like $\sqrt{2N_\tau}$ (no $v$ dependence at constant temperature). This can be done by assuming that any "exact" charge in $v$, for defining $P_v(Q)$, can get smeared by the same gaussian distribution with a certain width $\sigma$, representing the fluctuations. Finally we can include a very primitive, small short range interaction, telling us that instanton-antiinstantons annihilate if they get too close together ($< R$), by just shifting the class in these close encounters to class 0, when computing $f_n^v(Q)$.

In Fig.3 and Fig.4 we have included the best fit of our model to the data (obtained with the couplings chosen in last section, corresponding to a fine lattice). We find, using our model for a full volume of $V = 320 \times 6$ (solved up to 3 instanton-antiinstanton pairs), a good fit for both sub-volumes $v$, with the parameters in lattice units $\rho \approx .0013$, $\sigma \approx .27$, $L_I \approx 28$ and $R \approx 4$. The fact of having used only 3 pairs to calculate the fit, makes an overestimation for $\rho$ which should shift this value down to $\rho \approx .0011$ in the exact case, close to the value that we will get from the susceptibility.

In the following we will calculate the second moment of $P_v(Q)$, $<Q_v^2>$, which will be easier to calculate than the full $P$ and will give us directly $\rho$, for the case of non-interacting instantons.

The contribution $p^{I,A}(Q)$ of a single instanton (antiinstanton) in a volume $V$ with uniform distributed position to the total probability distribution of $Q$ measured in $v$ ($v \geq L_I$) includes the probabilities to be completely in $v$ ($p_{\text{in}} = \frac{v - L_I N_t}{V}$), completely out of $v$ ($p_{\text{out}} = \frac{V - L_I N_t - v}{V}$) and to lie on the surface ($p_{\text{surf}} = \frac{2 L_I N_t}{V}$), finally resulting in

$$p^{I,A}(Q) = p_{\text{out}} \delta(Q) + p_{\text{in}} \delta(Q \mp 1) + p_{\text{surf}} \,. \tag{14}$$

To this distribution of excitations we have added independent gaussian quantum fluctuations, denoted by $p^G(Q)$ of the topological charge with width $\sigma$, which is in general a function of $v$.

The expectation value of $Q_v^2$ can be calculated exactly to give (using $p^I(Q) = p^A(-Q)$)

$$\begin{aligned}
< Q_v^2 > &= \sum_{n=0}^{\infty} p_n(\rho V) \int_{-\infty}^{\infty} dQ \int_0^1 \prod_{j=1}^{2n} dQ_j \, (Q + \sum_{j=1}^{n} Q_j - Q_{n+j})^2 p^G(Q) \prod_{j=1}^{2n} p^I(Q_j) \\
&= \sigma^2 + 2(\frac{v}{V}(1-\frac{v}{V}) - \frac{L_I N_t}{3V}) \sum_n n p_n(\rho V) \, .
\end{aligned} \qquad (15)$$

The sum in the last expression may be approximated to $\sum_n n p_n(\rho V) \approx \rho V$ for $\rho V \gg 1$.

We did several Monte Carlo runs for volumes at various temperatures $T$, $V = 320 \times (4, 6, 8, 10, 12)$, and for various subvolumes $v = (60, 80, 120, 160) \times N_\tau$ in order to check eq.(15). We found indeed, for this range of $v$ for a fixed $N_\tau = 1/T$, that we could extract a fairly stable $\rho(N_\tau)$. There is a temperature dependence, as expected [24] from the squeezing of instantons at high $T$, obtaining respectively:
$\rho(4, 6, 8, 10, 12) = (\ 0.4(2),\ 1.1(2),\ 1.4(2),\ 1.4(1),\ 1.5(2)) \times 10^{-3}$. This instanton density $\rho$ grows fast as we decrease the temperature and starts stabilizing at low temperatures as wanted. Note that the density for $N_\tau = 6$ obtained here agrees well with the one extracted from $P_v(Q)$.

Since topological charge in a volume $v$ may be expressed as an integral over the boundary of $v$, statistically independent fluctuations within $v$ do not affect $Q^2$ and hence $\sigma^2$ will be proportional to the surface of $v$ for sufficiently large $v$. In our case $\partial v$ consists of two rings with a fixed length running around the torus. If the separation $v_x$ of the two lines is sufficiently large the fluctuation parts of the corresponding loop integrals, denoted by $< K(x) >$ and $< K(x + v_x) >$, are independent and $\sigma$ does not depend on $v_x$. This formal result is confirmed by our numerical calculations.

In contrast, the instanton contribution to $< Q_v^2 >$ has (apart from finite size effects, which is only effective on the surface) a different statistical behavior because the instanton parts of the loop integrals can never be independent. (If they were $< Q_v^2 >$ and thus the action density would be infinite. In fact, formally speaking, the loop integrals are maximally correlated for any $v_x$ since we have for the covariance

$$\frac{< K(x)K(x + v_x) >}{\sqrt{< K(x)^2 >< K(x + v_x)^2 >}} = 1 - \frac{1}{2} \frac{< Q^2(v) >}{\sqrt{< K(x)^2 >< K(x + v_x)^2 >}} = 1 \, ,$$

which is merely a consequence of the fact that the number of vacua is infinite.)

The different statistical behavior of the instanton and the fluctuation contribution can be used to distinguish the two. We believe that this is in general true as long as you can express topological charge in a volume $v$ in terms of the boundary of $v$.

## 5 Beyond optimal conditions

### 5.1 The role of dislocations

So far the simulations were done for parameters for which in practice any configuration was absolutely free of dislocations. Dislocations can occur when after an update a plaquette angle $\theta_n$ starts fluctuating around $2\pi n$ with a dislocation charge $n \neq 0$. In general due to the modulus in eq.(7) for $N$ dislocations with charges $n_i$ the topological charge of the whole lattice is

$$Q^{\text{tot}} = -\sum_{i=1}^{N} n_i .\qquad(16)$$

| n | 0 | 1 | 2 | 3 |
|---|---|---|---|---|
| with dislocations | 1. | .51±.04 | .11±.015 | .005±.002 |
| without dislocations | 1. | .53±.02 | .10±.01 | .007±.003 |

Table 1: Relative weights of sectors with charge $n$

A proper choice of parameters alone is not sufficient to avoid dislocations. A hot start of the Monte-Carlo simulation with random link variables will always produce dislocations being more and more stable the closer one comes to the continuum limit.

Since you can move a dislocation by adding an integer multiple of $2\pi$ to an angle of a link variable, which affects neither the action nor the topological density, it is clear that the system does not 'know' about the position of a dislocation. Furthermore, since dislocations of opposite sign (and equal magnitude) can be moved through the lattice to cancel each other, the only relevant quantity can be the net dislocation charge, which merely forces the 'real' total topological charge to assume the value $Q^{\text{tot}}$ and thus effectively acts like a different boundary condition (costing some action). When doing a Monte-Carlo simulation the configurations may now be divided into a discrete set of sectors with fixed $Q^{\text{tot}}$. Now we will address the question of whether a lattice of volume $v$ with inclusion of dislocations, behaves like a subsystem (with the same volume) of a larger lattice $V$ without dislocations.

When doing a careful analysis we must first of all make sure that we are in a parameter range where the jumps to different sectors are rare, for otherwise these jumps itself would spoil thermal equilibrium of each individual sector. We achieved this by choosing $\beta = 6.3$ without changing $\lambda$ and $\kappa$. For this $\beta$ a jump occurs approximately every 300 iterations.

To check numerically whether only the net dislocation charge can affect the system we partly modified the lagrangian to prohibit occurences of dislocations without changing the continuum limit. To be specific, we added a term $c \cdot \Theta(|\theta_n| - \pi)$, with c being a large constant, to the lagrangian density. We now compared results with the modified lagrangian, to the $Q^{\text{tot}} = 0$ sector of the original lagrangian, choosing all parameters identical and we found that all observed quantities including the probability distribution of $Q$ come out identically.

Furthermore, though the absolute amount of dislocation charges always increases (like a random walk), the distribution of the net dislocation charge, denoted by $\tilde{P}(n)$, converges with Monte-Carlo time and is (relative to the former) highly peaked around $Q^{\text{tot}} = 0$. We compared the distribution of $Q^{\text{tot}}$, obtained in a $40 \times 4$ lattice, with results from the modified version where we have measured the distribution $P_v(Q)$ in a $40 \times 4$ subvolume of a $320 \times 4$ lattice. For the latter the separation into sectors of charge $n$ was simply done by integrating $P_v(Q)$ over intervals $[n - \frac{1}{2}, n + \frac{1}{2}]$.

The resulting probability distribution of the sectors is shown in table 1. To make the relative weights of the sectors obvious we have normalized both distributions to $\tilde{P}(0) = 1$. The agreement of the two methods, within the errors in Table 1, confirm our hypothesis that dislocations simulate a larger volume in which the lattice can be thought to be embedded. This especially means that the dislocations seem to come with the right entropy. Therefore, the more usual way to measure the topological charge, based on using the dislocations to get non-zero charge seems fine, at least when the physical instantons are much larger than the lattice spacing. Clearly when the instantons are of the size of $a$, their action gets distorted and small, as they are not distinguishable from the dislocations anymore. Then the picture above loses meaning, but we were hopelessly out of the continuum scaling region.

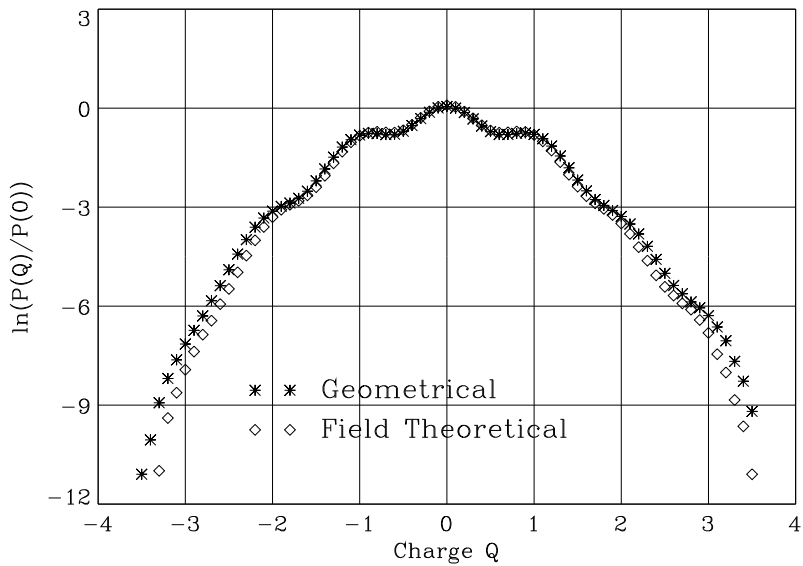

Figure 5: Both definitions give similar curves for fine lattices, the field theoretical one losing some information at large $Q$.

## 5.2 The field theoretical definition of topological charge

For non-abelian theories it is difficult to implement the geometrical definition of topological charge [19, 20, 10]. A far more simple definition is given by the field theoretical one [21] in which the topological density (in general $tr(FF^*)$) is expressed in terms of the plaquettes with the correct continuum limit.

The most severe problem with the field theoretical definition in eq.(5) is the (in contrast to dislocations uncontrollable) violation of eq.(4). Fluctuations within a volume $v$ do not cancel exactly, as happens with the geometrical definition in eq.(6). As a consequence, the fluctuation part of $Q$ measured in $v$ will, like the instantons themselves, have a contribution proportional to $\sqrt{v}$. The different statistical behavior of instanton and fluctuation contributions to $Q$ can not be used to distinguish the two and the deviation from the gaussian distribution of $Q$, the peaks at integer values, will be washed out, the larger one chooses $v$.

However, the difference of the two definitions of topological charge $\sum_n (\sin(\theta_n) - [\theta_n])$ is only of order $[\theta_n]^3$ and there is hope that for small fluctuations the peaks are still visible. In our case this is in fact true for moderate $v$ as is shown in Fig.5. Here we see good agreement of the probability distributions. In the non-abelian case though, one expects a poorer agreement [21], as the deviations from the geometrical case do not decrease for finer lattices.

# 6 Conclusion

We have seen that, at least in the two dimensional case, one can obtain extended physical instantons in a periodic box without any lattice artifacts. These instantons can be identified and their mean density $\rho$ and size measured, by considering the charge $Q_v$ in sub-volumes $v$ of the full lattice. The probability distribution to get some topological charge in $v$, $P_v(Q)$, can be obtained by lattice simulations and gives clear enhancements at integer $Q$. A simple model of almost non-interacting extended (anti)instantons fits nicely the Monte Carlo results. From a volume analysis of the susceptibility $<Q_v^2>$ we can easily extract the mean instanton density $\rho$ and find that it grows as we decrease the temperature until reaching some value at zero $T$, as expected.

Furthermore, we have seen that if one changes the couplings so as to allow some dislocations, still for fine enough lattices, the system acts as if we had changing boundary conditions for some number of physical instantons, having practically the same probability distribution as in the case above for instantons in a sub-volume $v$ of the same size and very large $V$. This is the more usual way to get topological charge distributions, but it was interesting to check that it really gives similar answers to the proper distribution without lattice artifacts.

This whole idea of measuring the topological charge on subvolumes of the periodic lattice, with $v$ much larger than the ultraviolet fluctuations correlation lenghts and even larger than the extended instantons and smaller than $V/2$, could be similarly tried out for non-abelian theories in higher dimensions. Clearly, the geometrical definition for topological charge is preferable at least as control. One would need though, fairly large average instantons sizes, in order to avoid dislocations. The larger dispersion in the sizes of instantons, compared to our 2-D case, might be the biggest problem. Improved actions certainly would help. The lattice would only need to be elongated in one direction, in order to decrease the fluctuations. Finally, the field theoretical definition could be useful at intermediate volumes for fine enough lattices.

### Acknowledgments

We would like to thank L. Polley for several interesting conversations related to the choice of boundary conditions. We thank the computing centre RRZN at Hannover and the local facilities for making the simulations possible.

# References

[1] A. Belavin and A. Polyakov JETP Lett. 22, 245 (1975)

[2] G. 't Hooft, Phys. Rev. Lett. 37, 8 (1976)

[3] A. Ringwald, Phys. Lett. B201, 510 (1988)

[4] V. Kuzmin, V. Rubakov, M. Shaposhnikov, Phys. Lett. 155B, 36 (1985)

[5] G. 't Hooft, Phys. Rev. D14, 3432 (1986)

[6] G. 't Hooft, Proc. of the EPS International Conf. on High Energy Physics, Palermo, 1975, ed. A. Zichichi (Editrice Compositori, Bologna, 1976)

[7] S. Mandelstam, Phys. Rep. C23 245 (1976) 245

[8] C. Callan, R. Dashen and D. Gross, Phys. Lett. 63B, 334 (1976)

[9] For more refs. on semiclassical methods see textbook: R. Rajaraman, Solitons and Instantons, North-Holland Phys. Pub., 1982.

[10] I. Fox, J. Gilchrist, M. Laursen and G. Schierholz, Phys. Rev. Lett. 54, 749 (1985)

[11] M. Göckeler, A. Kronefeld, M. Laursen, G. Schierholz and U.J. Wiese, Phys. Lett. B233,192 (1989)

[12] M. Chu, J. Grandy, S. Huang and J. Negele, Nucl. Phys. B34 (Proc. Suppl.), 170 (1994)

[13] D. Grigoriev, V. Rubakov and M. Shaposhnikov, Phys. Lett. B 261, 172 (1989)

[14] S. Grunewald, E. Ilgenfritz and M. Müller-Preussker, Z. Phys. C 33, 561 (1987)

[15] J. Hoek, M. Teper and J. Waterhouse, Nucl. Phys. B288, 589 (1987)

[16] M. Garcia, A. Gonzalez-Arrollo, J. Snippe and P. van Baal, Nucl. Phys. B34 (Proc. Suppl.), 222 (1994)

[17] P. de Forcrand, M. Garcia and I. Stamatescu, preprint IPS-95-21, Jul. 1995, Hep-lat/9509064.

[18] C. Panagiotakopoulos, Nucl. Phys. B251, 61 (1985)

[19] M. Lüscher, Commun. Math. Phys. 85, 39 (1982)

[20] P. van Baal, Commun. Math. Phys 85, 529 (1982)

[21] P. di Vecchia, K. Fabricius, G. Rossi and G. Veneziano, Nucl. Phys. B192, 392 (1981)

[22] H.Evertz, K.Jansen, J.Jersak, C.Lang and T.Neuhaus, Nucl. Phys. B285, 590 (1987)

[23] B. Berg and C. Panagiotakopoulos, Phys. Rev. Lett. 52,94 (1984)

[24] D. Gross, R. Pisarski and L. Yaffe, Rev. of Mod. Phys. 53, 43 (1981)